\documentclass[prl,aps,showpacs,onecolumn,floatfix]{revtex4}
\usepackage{epsfig} \usepackage{graphics} \usepackage{bm}
\usepackage{amssymb}
\usepackage{graphicx}
\begin{document}

\preprint{Lebed-2020}

\title{Restoration of superconductivity in
high magnetic fields in UTe$_2$}

\author{Andrei G. Lebed}

\address{Department of Physics, University of Arizona,\\ 1118 E. 4th Street
Tucson, Arizona 85721, USA \\lebed@physics.arizona.edu}

\address{L.D. Landau Institute for
Theoretical Physics,\\ RAS, 2 Kosygina
Street, Moscow 117334, Russia \\
}

\begin{abstract}
It was theoretically predicted more that 20 years ago [A.G. Lebed
and K. Yamaji, {\it Phys. Rev. Lett.} {\bf80}, 2697 (1998)] that a
triplet quasi-two-dimensional (Q2D) superconductor could restore
its superconducting state in parallel magnetic fields, which are
higher than its upper critical magnetic field, $H > H_{c2}(0)$. It
is very likely that recently such phenomenon has been
experimentally discovered in the Q2D superconductor UTe$_2$ by
Nicholas Butch, Sheng Ran and their colleagues and has been
confirmed by Japanese-French team. We review our previous
theoretical results, using such a general method that it describes
the reentrant superconductivity in the above mentioned compound as
well as will hopefully describe the similar phenomena, which can
be discovered in other Q2D superconductors.
\end{abstract}


\maketitle

\section{1. Introduction}
The so-called reentrant superconductivity phenomenon,
experimentally observed in quasi-two-dimensional (Q2D) organic
superconductor $\lambda$-(BETS)$_2$FeCl$_4$ [1] as well as in
ferromagnetic superconductors URhGe [2,3] and UCoGe [4], have been
recently intensively studied both experimentally and
theoretically. In the case of the above mentioned organic
superconductor, the high-field superconducting phase has been
prescribed to Jaccarino-Peter effect [5], whereas the physical
origin of the reentrant phase in the ferromagnetic superconductors
was prescribed to the existence of ferromagnetic fluctuations
[6,7]. On the other hand, for layered Q1D [8-10] and for isotropic
within the layers Q2D triplet superconductors [11], many years
ago, there was suggested effect of reentrant superconductivity in
a parallel magnetic field. It was later confirmed in Refs.
[12-15]. Very recently, superconductivity and the reentrant
superconductivity have been discovered [16-19] in the
non-ferromagnetic Q2D [15,20] superconductor UTe$_2$. As was
stressed in Ref. [15], the above mentioned reentrant
superconductivity cannot be due to the ferromagnetic fluctuations
and are likely due to the effect of two-dimensionalization of
electron spectrum first theoretically predicted in Refs.[8,11].

\section{2. Goal}
Our goal is to review Refs.[8,11,12], using the general method
[12], that describes well the case of the Q2D superconductor
UTe$_2$ [20]. We hope that it would describe also possible
discoveries of the reentrant superconductivity, which may be done
in the future in different Q2D and Q1D materials. In other words,
we show that the reentrant superconductivity [8,11,12] appears in
Q2D and layered Q1D superconductors due to two-dimensionalization
of electron spectrum for arbitrary in-plane shapes of electron
spectra and arbitrary triplet equal-spin in-plane superconducting
electron interactions. It is important that our approach is
qualitatively also applied to two-band superconductivity, which
may exist in UTe$_2$ [20].

\section{3. Restoration of superconductivity in a general Q2D case}
In this section, we consider a general Q2D case [12], where
in-plane electron spectrum has an arbitrary shape and in-plane
electron-electron interactions are of a general form and promote a
triplet pairing, which is not sensitive to the Pauli
spin-splitting effects against superconductivity.

\subsection{3.1. Qualitative description of a general Q2D case}
In this subsection, we suggest strong qualitative arguments why
superconductivity restores in a Q2D triplet superconductor at very
high magnetic fields, $H > H_{c2}$. We consider a layered
superconductor with the following Q2D electron spectrum in a
metallic phase:
\begin{equation}
\epsilon({\bf p})= \epsilon_{\parallel} (p_x,p_y) + 2 t_{\perp}
\cos(p_z d), \ \ \  t_{\perp} \ll \epsilon_F,
\end{equation}
where arbitrary in-plane energy, $\epsilon_{\parallel} (p_x,p_y)$,
corresponds to closed Fermi surface (FS), $t_{\perp}$ is the
integral of overlapping of electron wave functions in a
perpendicular to the conducting planes direction, $d$ is a
distance between the conducting layers, and $\epsilon_F$ is the
Fermi energy. For the further development, it is convenient to
linearize Q2D electron spectrum (1) near the FS:
\begin{equation}
\epsilon^{\pm}({\bf p})-\epsilon_F = \pm |v_x(p_y)|[p_x \mp
|p_x(p_y)|] + 2 t_{\perp} \cos(p_z d),
\end{equation}
where $v_x(p_y)$ is $x$-component of the Fermi velocity on the FS,
$p_x(p_y)$ satisfies the following condition:
\begin{equation}
\epsilon_{\parallel}[p_x(p_y),p_y] = \epsilon_F,
\end{equation}
+(-) stands for $p_x(p_y)>0$[$p_x(p_y)<0$].

In a magnetic field,
\begin{equation}
{\bf H} = (0,H,0) , \ \ \ {\bf A} = (0,0,-Hx),
\end{equation}
electron quasiclassical motion on the FS occurs due to the
following $z$-component of the Lorenz force:
\begin{equation}
\frac{d p_z}{dt} = \frac{e}{c} v_x(p_y)H.
\end{equation}
Taking into account that
\begin{equation}
v_z(p_z) = \frac{2 t_{\perp} \cos(p_z d)}{d p_z} = - 2 t_{\perp} d
\sin(p_z d) ,
\end{equation}
we find that the electron motion between the conducting planes is
a trajectory oscillating in time,
\begin{equation}
z(t) = l_{\perp}(H, p_y) \ \cos[\omega_c(H, p_y)t].
\end{equation}
with the frequency and amplitude being:
\begin{equation}
 \omega_c(H, p_y) = \frac{e |v_x(p_y)| H d}{c}, \ \ \ l_{\perp}(H,
 p_y)=d \frac{2 t_{\perp}}{ \omega_c(H, p_y)}.
\end{equation}
\begin{figure}[th]
\centerline{\includegraphics[width=12cm]{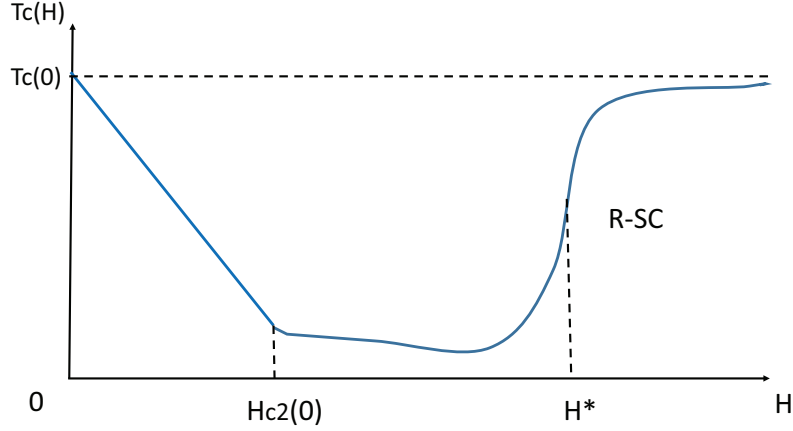}} \vspace*{8pt}
\caption{A schematic illustration of the restoration
superconductivity phenomenon. As seen from the figure,
$T_c(H>>H^*) \approx T_c(0)$, where the critical magnetic field
$H^*$ is given by Eq.(9). R-SC stands for the reentrant
superconducting phase. \label{f1}}
\end{figure}

From Eqs.(7) and (8), it directly follows that the amplitudes of
electron motion between the conducting planes in a magnetic field
decrease with the increasing magnetic field (4). In very high
magnetic fields, the electron amplitudes become less than the
distance between the planes, $d$, for the majority of electrons.
It happens when
\begin{equation}
H \geq H^* = \frac{2 t_{\perp} c}{e v_F d},
\end{equation}
where $v_F$ is a characteristic velocity of in-plane electron
motion. In this case, the destructive Meissner currents
perpendicular to the planes become small and superconducting state
has to restore (see Fig.1). This is a qualitative explanation of
the two-dimensionalization phenomena of electron spectrum which
lead to the restoration of superconductivity in high magnetic
fields [8-15]. We pay attention that the restoration of
superconductivity has to occur for any in-plane anisotropic
electron spectrum (1) and for any equal spin in-plane
electron-electron interactions. Generalization of the
two-dinsionalization phenomenon for two-band superconductivity is
straightforward.

\subsection{3.2. Quantitative description of a general Q2D case}
Here, we use the Green's functions method to quantitatively
establish the two-dimensionalization phenomenon and the
restoration of superconductivity in high magnetic fields in a
triplet Q2D superconductor with a general in-plane electron
spectrum and with general in-plane electron-electron
interactions[12,11]. In a magnetic field ${\bf H}
\parallel {\bf y}$ in the gauge (4), we can use for the electron
spectrum (2) the so-called Peierls substitution method,
\begin{equation}
p_x \rightarrow -i \frac{d}{dx}, \ \ \ p_z \rightarrow p_z
-\frac{e}{c} A_z = p_z+\frac{e}{c} Hx.
\end{equation}
In this case, we obtain the following Schr\"{o}dinger-like
equation for non-interacting electron wave functions in the
magnetic field (4):
\begin{eqnarray}
\biggl\{\pm |v_x(p_y)| \biggl[-i \frac{d}{dx} \mp
&|p_x(p_y)|\biggl] + 2 t_{\perp} \cos \biggl(p_z d + \frac
{eHdx}{c}\biggl) -2 \mu_BH \sigma \bigg\}
\nonumber\\
&\times\Psi_{\epsilon}(x,p_y,p_z; \sigma)= \epsilon
\Psi_{\epsilon} (x,p_y,p_z; \sigma)
\end{eqnarray}
The solutions of Eq.(11) for the electron wave functions are:
\begin{eqnarray}
&\Psi_{\epsilon}(x,p_y,p_z; \sigma) =  \frac{1}{\sqrt{|v_x(p_y)|}}
\exp \biggl[\frac{\pm i \epsilon x}{|v_x(p_y)|} \biggl]\exp [\pm i
|p_x(p_y)|x]
\nonumber\\
&\times\exp \biggl[\frac{\pm 2 i \mu_BH \sigma x}{|v_x(p_y)|}
\biggl] \exp\biggl\{\frac{\mp i \lambda(p_y)}{2} \biggl[
 \sin \biggl(p_z d + \frac{e H d x}{c} \biggl) \biggl] \bigg\} ,
\end{eqnarray}
where $\mu_B$ is the Bohr magneton, $\sigma = \pm \frac{1}{2}$ is
y-component of the electron spin; $\lambda(p_y)=4t_{\perp} c /e
|v_x(p_y)|H d $.
 Let us define superconducting
transition temperature in the magnetic field (4). To this end, it
is convenient to introduce equation for temperature (Matsubara's)
Green's function [21]. In according with Eq.(11) and Ref.[21], the
Green's functions, $G^{\pm}_{i \omega_n}(x,x_1;p_y,p_z; \sigma)$,
obey the following equation:
\begin{eqnarray}
\biggl\{-i\omega_n \pm |v_x(p_y)| \biggl[-i \frac{d}{dx} \mp
&|p_x(p_y)|\biggl] + 2 t_{\perp} \cos \biggl(p_z d + \frac
{eHdx}{c}\biggl) -2 \mu_BH \sigma \bigg\}
\nonumber\\
&\times G^{\pm}_{i \omega_n}(x,x_1;p_y,p_z; \sigma)=
\delta(x-x_1),
\end{eqnarray}
where $\delta(x-x_1)$ is the Dirac delta-function. It is important
that Eq.(13) can be analytically solved:
\begin{eqnarray}
&G^{\pm}_{i \omega_n}(x,x_1;p_y,p_z; \sigma) = -i
\frac{sgn(\omega_n)}{|v_x(p_y)|} \exp \biggl[\frac{\mp \omega_n
(x-x_1)}{|v_x(p_y)|} \biggl]\exp [\pm i |p_x(p_y)|(x-x_1)]
\nonumber\\
&\times \exp \biggl[\frac{\pm 2 i \mu_BH \sigma
(x-x_1)}{|v_x(p_y)|} \biggl]
\nonumber\\
&\times \exp\biggl\{\frac{\mp i \lambda(p_y)}{2} \biggl[
 \sin \biggl(p_z d + \frac {e H d x}{c}\biggl)  - \sin \biggl(p_z d +
 \frac{e H d x_1}{c}\biggl) \biggl]  \bigg\} ,
\end{eqnarray}
where $\mp \omega_n (x-x_1) <0$.

The so-called gap equation, determining the upper critical
magnetic field temperature dependence, $H_{c2}(T)$, is derived by
means of the Gor'kov's equations [22] for non-uniform
superconductivity [23,24]. As a result, we obtain:
\begin{eqnarray}
&\Delta(p_x,p_y;x) = \int dp^1_y \int^{\infty}_{|x-x_1| >
|v_x(p_y^1)|/\Omega} \frac{2 \pi T dx_1}{v_x^2(p_y^1)
\sinh\biggl[\frac{2 \pi T|x-x_1|}{|v_x(p_y^1)|} \biggl]}
\nonumber\\
&\times J_0 \biggl\{ 2 \lambda(p^1_y) \sin \biggl[
\frac{eHd(x-x_1)}{c}\biggl] \sin \biggl[ \frac{eHd(x+x_1)}{c}
\bigg]\biggl\}
\nonumber\\
&\times \cos \biggl[ \frac{2 \mu_B(1-S)H(x-x_1)}{|v_x(p_y^1|)}
\biggl]
\nonumber\\
&\times \{U[p_x,p_y;|p^1_x (p_y^1)|,p_y^1] \Delta[|p^1_x
(p_y^1)|,p_y^1;x_1]
\nonumber\\
&+U[p_x,p_y;-|p^1_x (p_y^1)|,p_y^1] \Delta[-|p^1_x
(p_y^1)|,p_y^1;x_1]\} ,
\end{eqnarray}
where the order parameter, $\Delta(p_x,p_y;x)$, depends on the
position of a center of mass of the Cooper pair, $x$, as well as
on the position on the FS [$p_x$ and $p_y$ satisfy the following
condition: $\epsilon_{\parallel}(p_x,p_y) =\epsilon_F$];
$U[p_x,p_y;p_x(p_y^1),p_y^1)]$ is a matrix element of the
electron-electron interactions; $S=0,1$ is the total spin of the
Cooper pair; $\Omega$ is a cutoff energy.[Note that, in Eq.(15),
the Bessel function, $J_0(...)$, describes the orbital effects
against superconductivity in a magnetic field, whereas $\cos[...]$
describes the destructive Pauli spin-splitting paramagnetic
effects]. Below, we consider the case of triplet equal-spin
pairing, therefore, $S=1$ in Eq.(15) and, thus, the Pauli
paramagnetic effects against superconductivity are absent.

Thus, we can rewrite Eq.(15) in the following form:
\begin{eqnarray}
&\Delta(p_x,p_y;x) = \int dp^1_y \int^{\infty}_{|x-x_1| >
|v_x(p_y^1)|/\Omega} \frac{2 \pi T dx_1}{v_x^2(p_y^1)
\sinh\biggl[\frac{2 \pi T|x-x_1|}{|v_x(p_y^1)|} \biggl]}
\nonumber\\
&\times J_0 \biggl\{ 2 \lambda(p^1_y) \sin \biggl[
\frac{eHd(x-x_1)}{c} \biggl] \sin \biggl[ \frac{eHd(x+x_1)}{c}
\bigg]\biggl\}
\nonumber\\
&\times \{U[p_x,p_y;|p^1_x (p_y^1)|,p_y^1] \Delta[|p^1_x
(p_y^1)|,p_y^1;x_1]
\nonumber\\
&+U[p_x,p_y;-|p^1_x (p_y^1)|,p_y^1] \Delta[-|p^1_x
(p_y^1)|,p_y^1;x_1]\} .
\end{eqnarray}
Then, by means of relationships,
\begin{equation}
\frac{d p_y}{v_x(p_y)}= \frac{d p_l}{v_{\perp}(p_l)}, \ \ \ dp_l^2
= dp_x^2 + dp_y^2, \ \ v^2_{\perp} (p_l) = v_x^2 (p_l) + v_y^2
(p_l),
\end{equation}
we can express Eq.(16) as
\begin{eqnarray}
&\Delta(p_l;x) = \oint \frac{dp^1_l}{v_{\perp}(p^1_l)}
\int^{\infty}_{|x-x_1|
> |v_x(p_l^1)|/\Omega} \frac{2 \pi T dx_1}{|v_x(p_l^1)|
\sinh\biggl[\frac{2 \pi T|x-x_1|}{|v_x(p_l^1)|} \biggl]}
\nonumber\\
&\times J_0 \biggl\{ 2 \lambda(p^1_l) \sin \biggl[
\frac{eHd(x-x_1)}{c} \bigg] \sin \biggl[ \frac{eHd(x+x_1)}{c}
\bigg]\biggl\}
\nonumber\\
&\times U[p_l;p_l^1] \Delta[p_l,p_l^1;x_1],
\end{eqnarray}
where integration in Eq.(18) is performed over the FS contour.

Let us introduce new variable $z$,
\begin{equation}
x_1 = x + z |v_x(p^1_l)|/v_F, \ \ \ v_F = <|v_{x}(p_l)|>_{p_l} ,
\end{equation}
$<>_{p_l}$ is an average value over the FS. In this new variable
of integration the gap equation (18), can be rewritten in the
following more convenient way:
\begin{eqnarray}
&\Delta(p_l;x) = \oint \frac{dp^1_l}{v_{\perp}(p^1_l)}
\int^{\infty}_{|z|
> v_F/\Omega} \frac{2 \pi T dz}{v_F
\sinh \biggl[\frac{2 \pi T|z|}{v_F} \biggl]}
\nonumber\\
&\times J_0 \biggl\{ 2 \lambda(p_l^1) \sin \biggl[ \frac{e d H z
|v_x(p_l^1|)}{c} \bigg] \sin \biggl[ \frac{e d H
(2x+z|v_x(p_l^1)|/v_F)}{c} \bigg]\biggl\}
\nonumber\\
&\times U[p_l;p_l^1] \Delta[p_l,p_l^1; x + z |v_x(p^1_l)|/v_F],
\end{eqnarray}
The effect of the two-dimensionalization of the Q2D electron
spectrum (1) and the restoration of superconductivity phenomenon
in a magnetic field are directly seen from Eq. (20), where
\begin{equation}
\lambda(p^1_l)=\frac{2 |l_{\perp}(p^1_l)|}{d}
\end{equation}
is a dimensionless magnitude of electron trajectory in the
perpendicular to the planes direction, expressed in terms of the
inter-plane distance, $d$. If $H \geq H^*$, where the critical
field $H^*$ is given by Eq.(9), then $|\lambda(p^1_l)| \leq 1$ for
the significant part of electrons on the Q2D FS. In this case, the
Bessel function $J_0(...) \approx 1$ in Eq. (20) and, therefore,
Eq. (20) has the same solutions as without the magnetic field (4):
\begin{equation}
\Delta(p_l) = \oint \frac{dp^1_l}{v_{\perp}(p^1_l)}
\int^{\infty}_{|z|
> v_F/\Omega} \frac{2 \pi T dx_1}{v_F
\sinh \biggl[\frac{2 \pi T|z|}{v_F} \biggl]} U[p_l;p_l^1]
\Delta(p_l,p_l^1).
\end{equation}
For this reason superconductivity restores in the triplet case at
$H \geq H^*$ with the same transition temperature, as it has in
zero magnetic field (see Fig.1):
\begin{equation}
T_c(H \gg H^*) \approx T_c(0).
\end{equation}
As we mentioned in the previous subsection, the physical meaning
of the restoration of superconductivity is that electrons are
almost localized on the conducting planes and, therefore, the
destructive Meissner currents are significantly suppressed at
$|\lambda(p^1_l)| \leq 1$. In the case of s(d)-wave
superconducting pairing [i.e., at $S=0$ in the Eq.(15)], the above
described phenomenon creates opportunity [8,11] for
superconductivity to exist at $H>H_{p}(0)$ in the form of the
so-called Larkin-Ovchinnikov-Fulde-Ferrel (LOFF) phase [25,26],
where $H_p$ is the so-called paramagnetic limit [27,28].

\section{4. Restoration of superconductivity in an in-plane isotropic Q2D case}

Let us consider an important limiting case with in-plane isotropic
electron spectrum and the simplest in-plane triplet
superconducting electron-electron interactions [11].

\subsection{4.1. Qualitative description of an in-plane isotropic Q2D case}
As usual, we start from qualitative description of the
two-dimensionalization of in-plane isotropic Q2D electron spectrum
and its consequence - the phenomenon of the restoration of
superconductivity in high magnetic fields, $H > H_{c2}$. Instead
of arbitrary electron spectrum, here we consider a layered
superconductor with the following in-plane isotropic Q2D electron
spectrum in a metallic phase:
\begin{equation}
\epsilon({\bf p})= \frac{p^2_x+p_y^2}{2m} + 2 t_{\perp} \cos(p_z
d), \ \ \ t_{\perp} \ll \epsilon_F,
\end{equation}
where isotropic in-plane energy corresponds to the in-plane closed
FS, $t_{\perp}$ is the integral of overlapping of electron wave
functions in a perpendicular to the conducting planes direction,
$d$ is a distance between the conducting layers, and $\epsilon_F$
is the Fermi energy. For calculation of the quasi-classical
electron trajectories, it is convenient, as usual, to linearize
Q2D electron spectrum (24) near the FS:
\begin{equation}
\epsilon^{\pm}({\bf p})-\epsilon_F = \pm v_F |\sin \phi|[p_x \mp
p_F |\sin \phi|] + 2 t_{\perp} \cos(p_z d),
\end{equation}
where we count the polar angle $\phi$ from y-axis. In Eq.(25),
$v_F$ is the Fermi velocity, $p_F =m v_F$ is the Fermi momentum;
$v_F sin \phi$ is $x$-component of the Fermi velocity, $p_F \sin
\phi$ is $x$-component of the Fermi momentum, which satisfies the
following condition:
\begin{equation}
\frac{p_F^2 \sin^2 \phi + p_F^2 \cos^2 \phi}{2m} =
\frac{p_F^2}{2m}= \epsilon_F.
\end{equation}

In the external magnetic field (4), electron motion on the FS is
due to the action of the following z-component of the Lorenz
force:
\begin{equation}
\frac{d p_z}{dt} = \frac{e}{c} v_F \sin \phi \ H.
\end{equation}
It is known that in the quasiclassical approximation
\begin{equation}
v_z(p_z) = \frac{2 t_{\perp} \cos(p_z d)}{d p_z} = - 2 t_{\perp} d
\sin(p_z d) .
\end{equation}
Therefore, we find that electron trajectories between the
conducting planes are the oscillating functions of time,
\begin{equation}
z(t,\phi) = l_{\perp}(H,\phi) \ \cos[\omega_c(H, \phi)t],
\end{equation}
with the frequency and amplitude being:
\begin{equation}
 \omega_c(H,\phi) = \frac{e v_F |\sin \phi|  H d}{c}, \ \ \ l_{\perp}(H,
 \phi)=\frac{2 t_{\perp}}{ \omega_c(H,\phi)}.
\end{equation}
As seen from Eqs. (29) and (30) [compare to Eqs. (7) and (8)], the
amplitudes of electron motion between the conducting planes in a
magnetic field decrease with an increasing magnetic field. In very
high magnetic fields (9), the electron amplitudes become less than
the distance between the planes, $d$, for the significant part of
electrons. In this case, the destructive Meissner currents
perpendicular to the planes become small and superconducting state
has to restore. This is a qualitative explanation of the
two-dimensionalization phenomenon of electron spectrum, which
leads to the restoration of superconductivity in high magnetic
fields [8-15]. We pay attention that, as shown in previous
section, the restoration of superconductivity has to occur for any
in-plane anisotropic electron spectrum (1) and for any equal spin
in-plane electron-electron interactions.

\subsection{4.2. Quantitative description for an in-plane isotropic Q2D case}

In the case of in-plane isotropic Q1D spectrum, in the gap
Eq.(15), it is convenient to introduce two polar angles, $\phi$
and $\phi_1$, which we count from $y$-axis. Then gap Eq.(15) can
be rewritten in more simple way [11]:
\begin{eqnarray}
&\Delta(\phi;x) = \int_0^{2 \pi} \frac{d \phi_1}{2 \pi}
U(\phi,\phi_1) \int^{\infty}_{|x-x_1|
> v_F |\sin \phi_1|/\Omega} \frac{2 \pi T dx_1}{v_F |\sin \phi_1|
\sinh\biggl[\frac{2 \pi T|x-x_1|}{v_F |\sin \phi_1|} \biggl]}
\nonumber\\
&\times J_0 \biggl\{ \frac{2 \lambda}{|\sin \phi_1|} \sin \biggl[
\frac{\omega_c (x-x_1)}{2v_F}\biggl] \sin \biggl[ \frac{\omega_c
(x+x_1)}{2v_F} \bigg]\biggl\}
\nonumber\\
&\times \cos \biggl[ \frac{2 \mu_B(1-S)H(x-x_1)}{v_F |\sin
\phi_1|} \biggl] \Delta(\phi_1,x_1),
\end{eqnarray}
where
\begin{equation}
\lambda = \frac{4t_{\perp}}{\omega_c}, \ \ \ \omega_c =
\frac{ev_FHd}{c}.
\end{equation}
In Eq.(31), the superconducting gap, $\Delta(\phi,x)$, depends on
the coordinate of a center of mass of the Cooper pair, $x$, as
well as on the position on the FS, where $\phi$ is the polar
angles between $y$-axes and two component vector ${\bf p} =
[p_x(p_y), p_y]$, where $p_x^2 (p_y)+p_y^2 = p^2_F$. In this
review, we consider the case, where electron-electron interactions
depend only on in-plane momenta. In this section, in contrast to
the previous one, we consider the following simplest case of
triplet equal spin pairing:
\begin{equation}
U(\phi,\phi_1)= g \ u(\phi) \ u(\phi_1).
\end{equation}
In this case, we can rewrite Eq.(33) in more simple form:
\begin{eqnarray}
&\Delta(x) = \frac{g}{2} \int_0^{2 \pi} \frac{d \phi_1}{2 \pi}
\int^{\infty}_{|x-x_1|
> v_F |\sin \phi_1|/\Omega} u^2(\phi_1) \frac{2 \pi T dx_1}{v_F |\sin
\phi_1| \sinh\biggl[\frac{2 \pi T|x-x_1|}{v_F |\sin \phi_1|}
\biggl]}
\nonumber\\
&\times J_0 \biggl\{ \frac{2 \lambda}{|\sin \phi_1|} \sin \biggl[
\frac{\omega_c(x-x_1)}{2v_F}\biggl] \sin \biggl[ \frac{\omega_c
(x+x_1)}{2v_F} \bigg]\biggl\} \Delta(x_1),
\end{eqnarray}
where the superconducting gaps in Eqs.(31) and (34) are
\begin{equation}
\Delta(\phi,x)= u(\phi) \Delta(x),
\end{equation}
$g$ is dimensionless constant of electron coupling. By introducing
the more appropriate variable,
\begin{equation}
x_1-x = z |\sin \phi_1|.
\end{equation}
\begin{eqnarray}
&\Delta(x) = \frac{g}{2} \int_0^{2\pi} \frac{d \phi_1}{2 \pi}
\int^{\infty}_{|z|
> v_F/\Omega} \frac{2 \pi T dz}{v_F
\sinh\biggl[\frac{2 \pi T|z|}{v_F} \biggl]} u^2(\phi_1)
\nonumber\\
&\times J_0 \biggl\{ \frac{2 \lambda}{\sin \phi_1} \sin \biggl[
\frac{\omega_c z |\sin \phi_1|}{2v_F}\biggl] \sin \biggl[
\frac{\omega_c (2x+ z |\sin \phi_1)|}{2v_F} \bigg]\biggl\}
\Delta(x + z |\sin \phi_1|),
\end{eqnarray}
Since $\lambda$ is inversely proportional to a magnetic field [see
Eq.(32)], it is clear that in high magnetic fields, $H \geq H^*$
[see Eq.(9) and Fig.1], superconductivity has to restore with the
zero-field transition temperature [see Eq.(23)].

\section{5. Restoration of superconductivity in a layered Q1D case}

In this section, we consider an important limiting case of layered
Q1D superconductors with the simplest equal spin triplet
superconducting electron-electron interactions [8].

\subsection{5.1. Qualitative description of a layered Q1D case}

As in the previous sections, we begin our consideration from the
qualitative description of the phenomenon of the restoration of
superconductivity in high magnetic fields, $H > H_{c2}$. Instead
of a Q2D electron spectrum, here we consider a layered
superconductor with the following Q1D electron spectrum in a
metallic state:
\begin{equation}
\epsilon({\bf p})= 2t_a \cos(p_x a/2) + 2t_b \cos(p_y b)+ 2
t_{\perp} \cos(p_z d), \ \ \ t_{\perp} \ll t_b \ll t_a.
\end{equation}
This layered Q1D electron spectrum is realized in the so-called
Bechgaard salts - compounds with chemical formular (TMTSF)$_2$X,
where X=ClO$_4$, PF$_6$, AsF$_6$, etc. [We note that the compound
(TMTSF)PF$_6$ is considered as a candidate for a triplet electron
pairing.] The first term in Eq.(38) represents free-electron
motion along the chains ($t_a \simeq 2500 K$); whereas $t_b \simeq
250 K$ and $t_{\perp} \simeq 3-5 K$ are the overlapping integrals
of electron wave functions in the perpendicular to the conducting
chains directions. To calculate the quasi-classical electron
trajectories in the magnetic field (4), as usual, it is convenient
to linearize Q1D electron spectrum (38) near the Q1D FS's:
\begin{equation}
\epsilon^{\pm}({\bf p})-\epsilon_F = \pm v_F [p_x \mp p_F] + 2t_b
\cos(p_y b) + 2 t_{\perp} \cos(p_z d),
\end{equation}
where $v_F = t_a a \sin (p_F a/2) = t_a a / \sqrt{2}$ is the Fermi
velocity, $p_F = \pi/2a$ is the Fermi momentum, +(-) stands for
right(left) piece of the Q1D FS.

In the external magnetic field (4), electron motion on the
right(left) piece of the Q1D FS (39) satisfies the conditions:
\begin{equation}
\frac{d p_z}{dt} = \pm \frac{e}{c} v_F H.
\end{equation}
As usual, in the quasiclassical approximation
\begin{equation}
v_z(p_z) = \frac{2 t_{\perp} \cos(p_z d)}{d p_z} = - 2 t_{\perp} d
\sin(p_z d)
\end{equation}
and we find that electron trajectory between the conducting planes
is the following oscillating function in time,
\begin{equation}
z(t) = l_{\perp}(H) \ \cos[\omega_c(H)t],
\end{equation}
with the frequency and amplitude being:
\begin{equation}
 \omega_c(H) = \frac{e v_F H d}{c}, \ \
 \ l_{\perp}(H)= d \frac{2 t_{\perp}}{ \omega_c(H)}.
\end{equation}
As directly follows from Eqs. (42) and (43) [compare to Eqs. (7)
and (8)], the amplitude of electron motion between the conducting
planes in the magnetic field (4) becomes less than the inter-plane
distance for very high magnetic fields,
\begin{equation}
H \geq H^*,
\end{equation}
where $H^*$ is given by Eq.(9). As usual, in this case, the
destructive Meissner currents become small and superconductivity
has to restore with $T_c(H >> H^*) \approx T_c(0)$ (see Fig.1)

\subsection{5.2. Quantitative description of a layered Q1D case}

Below, we consider the simplest equal-spin triplet
electron-electron pairing in a Q1D case, where the superconducting
gap changes its sign on the different pieces of the FS. It
corresponds to the following electron-electron interactions:
\begin{equation}
U(p_x,p^1_x)= g \ sign(p_x) \ sign(p^1_x)
\end{equation}
and to the following superconducting gap:
\begin{equation}
\Delta(p_x;x)=  \ sign(p_x) \Delta(x).
\end{equation}
In this case, Eq.(15) can be rewritten in the more simple way:
\begin{eqnarray}
&\Delta(x) = \frac{g}{2} \int^{\infty}_{|x-x_1|
> v_F /\Omega} \frac{2 \pi T dx_1}{v_F
\sinh\biggl[\frac{2 \pi T|x-x_1|}{v_F} \biggl]}
\nonumber\\
&\times J_0 \biggl\{ 2 \lambda \sin \biggl[
\frac{\omega_c(x-x_1)}{2v_F}\biggl] \sin \biggl[ \frac{\omega_c
(x+x_1)}{2v_F} \bigg]\biggl\} \Delta(x_1),
\end{eqnarray}
where the parameters $\lambda$ and $\omega_c$ are defined by
Eq.(32). [Note that, in Eq.(47), the superconducting gap,
$\Delta(x)$, depends only on the coordinate of a center of mass of
the Cooper pair, $x$.] If we introduce, as usual, the more
convenient variable,
\begin{equation}
x_1-x = z,
\end{equation}
then
\begin{eqnarray}
&\Delta(x) = \frac{g}{2} \int^{\infty}_{|z|
> v_F/\Omega} \frac{2 \pi T dz}{v_F
\sinh\biggl[\frac{2 \pi T|z|}{v_F} \biggl]}
\nonumber\\
&\times J_0 \biggl\{ 2 \lambda \sin \biggl[ \frac{\omega_c
z}{2v_F}\biggl] \sin \biggl[ \frac{\omega_c (2x+ z)}{2v_F}
\bigg]\biggl\} \Delta(x + z),
\end{eqnarray}
Since $\lambda \sim \frac{1}{H}$, it is clear that, in high
magnetic fields, $H \geq H^*$, where $H^*$ is given by Eq.(9),
superconductivity is restored with the zero-field transition
temperature, $T_c(H>>H^*) \approx T_c(0)$ (see Fig.1).

\section{6. Conclusion}

In the review, we have discussed both qualitative and quantitative
pictures of the two-dimensionalizations effect in layered Q2D and
Q1D superconductors in a parallel magnetic field. We have
concentrated our attention on an important consequence of this
effect - the restoration of triplet superconductivity phenomenon,
first suggested by us in Q1D case in Ref.[8] and in Q2D case - in
Ref.[11] (see Fig.1). Our qualitative description is very general
one and is valid for arbitrary Q2D superconductors with arbitrary
2D electron-electron interactions, including two-band
superconductors. Our quantitative calculations are done for a
one-band arbitrary Q2D superconductor with arbitrary 2D
electron-electron interactions. We hope that the suggested
phenomenon describes not only the recently experimentally
discovered reentrant superconductivity in the triplet
superconductor UTe$_2$, but is useful for the future expeiments.

\section*{Acknowledgments}

The author is thankful to N.N. Bagmet (Lebed), N.P. Butch, and
Sheng Ran for the useful discussions.

\end{document}